\documentclass[useAMS,usenatbib]{mnras}
\usepackage{amsmath, amssymb, graphicx, hyperref, xcolor}

\newcommand{\lya}{Ly$\alpha$\hspace{1mm}}

\title[The circumgalactic medium in \lya: a new constraint on galactic outflow models]{The circumgalactic medium in Lyman-$\alpha$: a new constraint on galactic outflow models}
\author[Andrew S. Chung]
{\parbox{\textwidth}
{Andrew S. Chung$^{1, 4}$\thanks{E-mail: \texttt{andy@tensorlicio.us}},
Mark Dijkstra$^{2}$,
Benedetta Ciardi$^{1}$,
Koki Kakiichi$^{1,3}$,
Thorsten Naab$^{1}$
}\vspace{0.4cm}
\\$^{1}$Max Planck Institut f\"{u}r Astrophysik, Karl-Schwarzschild-Str. 1, 85741 Garching, Germany\\
$^{2}$Institute of Theoretical Astrophysics, University of Oslo, Postboks 1029, 0315 Oslo, Norway\\
$^{3}$Department of Physics and Astronomy, University College London, London, WC1E 6BT, UK\\
$^{4}$Tensorlicious, http://tensorlicio.us
}
\hypersetup{draft}
\begin{document}
\date{\today}

\pagerange{\pageref{firstpage}--\pageref{lastpage}} \pubyear{2017}
\maketitle
\label{firstpage}
\begin{abstract}
Galactic outflows are critical to our understanding of galaxy formation and evolution.
However the details of the underlying feedback process remain unclear.
We compare Ly$\alpha$ observations of the circumgalactic medium (CGM) of Lyman Break Galaxies (LBGs) with mock observations of their simulated CGM.
We use cosmological hydrodynamical `zoom-in' simulations of an LBG which contains strong, momentum-driven galactic outflows.
Simulation snapshots at $z=2.2$ and $z=2.65$ are used, corresponding to the available observational data.
The simulation is post-processed with the radiative transfer code \textsc{crash} to account for the impact of ionising photons on hydrogen gas surrounding the simulated LBG.  We generate mock absorption line maps for comparison with data derived from observed close galaxy-galaxy pairs.
We perform calculations of \lya photons scattering through the CGM with our newly developed Monte-Carlo code \textsc{slaf}, and compare to observations of diffuse \lya halos around LBGs.
Our fiducial galactic outflow model comes closer to reproducing currently observed characteristics of the CGM in \lya than a reference inefficient feedback model used for comparison. Nevertheless, our fiducial model still struggles to reproduce the observed data of the inner CGM (at impact parameter $b<30$kpc).
Our results suggest that galactic outflows affect Ly$\alpha$ absorption and emission around galaxies mostly at impact parameters $b<50$ kpc, while cold accretion flows dominate at larger distances. We discuss the implications of this result, and underline the potential constraining power of CGM observations - in emission and absorption - on galactic outflow models.
\end{abstract}

\section{Introduction}\label{introduction}
Understanding processes which govern galaxy formation and evolution is one of the challenges in modern cosmology.
In the standard $\Lambda$CDM cosmological scenario strong feedback is needed at both the low and high-mass ends of the galaxy mass function in order for models to match observations \citep{1991ApJ...379...52W}. `Feedback' typically refers to the complex processes through which star formation and accretion onto black holes deposit energy and momentum back into their surroundings. The details of feedback are not well understood (see \citealt{2005SSRv..116..625C} and \citealt{2016arXiv161206891N} for reviews on the topic). Because of its importance though, it is fundamental to study as many (potential) observational probes of this process as possible.

There is an increasing amount of data on the so-called `circum-galactic' medium (CGM), which has been defined as the region around galaxies out to a distance of $r\sim 300$ kpc and with a velocity offset from the galaxy's systemic redshift of up to $\Delta v \sim 300$ km s$^{-1}$ \citep{2012ApJ...750...67R}.
\citet{2010ApJ...717..289S} (hereafter S2010) note that the CGM provides a `laboratory in which the effects of galaxy formation and AGN accretion (e.g., radiative and hydrodynamical feedback and its recent history) can be measured on scales that are not accessible using direct observations of galaxies.
Indeed analyses of the COS-Halos Survey (\citealt{2014ApJ...792....8W}; \citealt{2017ApJ...837..169P}) show that the CGM is a dominant baryon reservoir on a galactic scale, and thus reinforces the notion that details of its composition, state, and dynamics may provide strong constraints on galaxy evolution models.

In this paper, we explore whether observations of the CGM in \lya can constrain galactic outflow models, by comparing simulations of the CGM with observations.

\citet{2011ApJ...736..160S} (hereafter S2011) found spatially extended \lya emission from the CGM of Lyman Break Galaxies (LBGs) by stacking \lya observations of 92 individual galaxies.
Similar stacking analyses have revealed (fainter/smaller) \lya halos around \lya selected galaxies (i.e. Lyman Alpha Emitters, LAEs) \citep[][]{2012MNRAS.425..878M, 2014MNRAS.442..110M, 2017ApJ...837..172X}\footnote{\citet{2013ApJ...776...75F} did not find \lya halos around $z\sim 2$ LAEs \citep[also see][]{2012AAS...21934006S}, and discuss that systematic uncertainties associated with stacking could reduce the statistical significance of previously reported detections.  \citet{2014MNRAS.442..110M} used larger samples of LAEs to confirm these systematic effects, but still obtained significant detections of \lya halos.}.
At low redshift (0.028 $< z <$ 0.18), the Lyman alpha Reference Sample \citep[][]{2013ApJ...765L..27H} likewise shows extended \lya emission relative to H$\alpha$ in imaging observations from the Hubble Space Telescope.

S2011 proposed that the diffuse halos arise from \lya photons, produced in star-forming regions, scattering off outflowing material as they escape the galaxy. The presence of these outflows was inferred from the ubiquitous blueshifted
low-ionisation absorption lines, and their interaction with \lya photons was inferred from the redshifted \lya emission lines  (S2010).
Here, star formation is both the source of the \lya photons (produced by recombination in the HII regions around young stars) and of the stellar feedback which drives the outflowing material. S2011 provided a simple analytic model for the scattering of \lya photons through the outflow, and showed that it is a good fit to their stacked observations. The radiative transfer of \lya photons was treated though in an extremely approximate fashion. 

Motivated by these models, \citet{2012MNRAS.424.1672D} performed a systematic study of \lya transfer through phenomenological models of spherically and cylindrically symmetric, large-scale, clumpy outflows.
In these models, clumps were exclusively outflowing, with a one-to-one correspondence between outflow velocity ($v$) and distance from the galaxy ($r$).
Following S2011, the velocity profile was inspired by `momentum-driven' wind models in which the outflow accelerates as $a \propto r^{-\alpha}$ ($\alpha \sim 1.5$). \citet{2012MNRAS.424.1672D} constrained the HI properties of their clumps by matching the galaxy-galaxy pair absorption line presented by S2010. While these models can simultaneously explain the presence of \lya halos and the amount of absorption, they also predict that  a non-negligible fraction of \lya photons did not scatter at all. These photons should be visible as a bright point source, which is absent in the data. As \citet{2012MNRAS.424.1672D} pointed out, this problem could be due to the absence of low-column density HI systems, and/or a consequence of the simplified velocity profile of the outflow in their models. 

In this work, we take a new complementary approach, and use cosmological hydrodynamical `zoom' simulations from \citet{2012ApJ...745...11G} to generate a model CGM of a LBG. These simulations contain strong galactic outflows which are also momentum-driven, and provide us with a complex CGM that may more closely reflect reality than previous models. Importantly, the simulations contain inflowing `cold streams' \citep[e.g.][]{2005MNRAS.363....2K,2009Natur.457..451D}, which can contribute significantly to the amount of absorption measured in the CGM \citep{2012MNRAS.421.2809V,2012MNRAS.424.2292G} and possibly to the emission \citep[e.g.][]{2009MNRAS.400.1109D,2012MNRAS.423..344R}.
While these simulations do not have the resolution to properly resolve the feedback processes and the kinematics of the cold gas, it is important to check how they compare to the available data on the CGM.

While there are a number of previous works which use simulations to study observational signatures of the CGM, our work distinguishes itself by simultaneously considering the CGM in \lya emission and absorption.
Previous works that used simulations have focussed either on emission \citep[e.g.][]{2009ApJ...696..853L,2010ApJ...725..633F,2011MNRAS.416.1723B,2012MNRAS.423..344R} or absorption \citep[e.g.][]{2012MNRAS.424.2292G,2013ApJ...765...89S}.
\citet{2012MNRAS.424.1672D} show that joint constraints from \lya in absorption and emission are much more powerful than either data set individually.
Observationally this is demonstrated by \citet{2013ApJ...766...58H}.

Finally, in contrast to previous studies that modelled the CGM in absorption, our simulations are post-processed with an ionising photon transfer code (\textsc{crash}, \citealt{2001MNRAS.324..381C}) and account for local sources of ionising radiation, which can be more important than the overall ionising background \citep[especially at close distances to the galaxy, see e.g.][]{2013ApJ...765...89S}.

This paper is laid out as follows: in \S~\ref{simulations} we describe the simulations, \S~\ref{results} presents the output of our pipeline and compares our results to observations, \S~\ref{conclusions} discusses the results and our conclusions\footnote{Throughout this work, we used the following cosmological parameters based on the 3-year WMAP results \citep{2007ApJS..170..377S}: $\Omega_\Lambda = 0.74$, $\Omega_m = 0.26$, $\Omega_b = 0.044$, $h=0.72$, $n=0.95$, and $\sigma_8=0.77$.}.

\section{Simulations}\label{simulations}

\subsection{Cosmological Hydrodynamical Simulations }
It is not currently computationally feasible to perform full hydrodynamic cosmological simulations with sufficient resolution to resolve the detailed gas dynamics of the CGM.
Therefore we use the cosmological `zoom-in' simulations from \citet{2012ApJ...745...11G}, which start with an N-body dark matter only simulation.
A region of space is cut out around a massive dark matter halo and re-simulated, adding baryons and hydrodynamic physics using a modified version of \textsc{gadget-2} (\citealt{2005MNRAS.364.1105S}, \citealt{2006MNRAS.373.1265O}, and \citealt{2008MNRAS.387..577O}).

\citet{2012ApJ...745...11G} use a modified \citet{2008MNRAS.387..577O} wind model, which implements momentum-driven winds powered by stellar feedback.
In brief, gas particles become star-forming particles as per the criteria described in \citet{2008MNRAS.383.1210S}, and then subsequently become wind particles which are stochastically kicked perpendicularly to the plane of the galaxy.
The strength of the kick given to a particular wind particle is given by $v_{wind} = \sigma (4 + 4.29\sqrt{f_L - 1})$, where $\sigma$ is the velocity dispersion of the galaxy, and $f_L$ is the luminosity factor stochastically chosen in the range [1.05-2].
With this model $v_{wind}$ is higher than that used in \citet{2008MNRAS.387..577O}, resulting in typical wind velocities of $\approx$ (400-700) km~s$^{-1}$ for the halos under consideration here.
This, combined with the fact that in this model wind particles are temporarily decoupled from hydrodynamics so that they initially encounter no drag, ensures that the wind particles escape the disk.
The mass-loading factor, which is the wind mass loss rate divided by the star formation rate, is typically $\eta \approx 4$.

It is worth noting that this wind model was used in a suite of simulations which reproduce the metallicity and ionisation of the intergalactic medium (IGM), the galaxy mass-metallicity relation, the high galactic gas fraction at high redshift ($z\gtrsim 2$), and the fact that galaxies contain a low fraction of cosmic baryons ($\sim 5-10\%$ at $z=0$ \citealt{2004ApJ...616..643F}).
In other words, despite its simplicity, this model simultaneously reproduces several observational constraints and scaling relations.

The hydrodynamic simulation, covering a region of $\approx$ 5~Mpc comoving,  has a mass resolution of $8 \times 10^{5} M_\odot$ for baryonic particles, and $5 \times 10^{6} M_\odot$ for dark matter particles.
The gravitational softening length of the baryonic particles is 200$h^{-1}$ pc comoving.
Since we ultimately want to compare to S2010 and S2011, we select galaxies with a similar stellar mass (i.e. $M_* \approx 10^{10.5} M_\odot$) and use the \textsc{gadget} snapshots at $z=2.2$ (absorption) and $z = 2.65$ (emission), which are equivalent to the mean redshifts of the observations in S2010 and S2011, respectively.
Further details of the specific galaxy under consideration in this paper (identified as $s396$) can be found in Table 1 of \citet{2012ApJ...745...11G}.
In brief, at the $z=2.2$ snapshot, $s396$ resides in a $1.5 \times 10^{12}M_{\odot}$ dark matter halo, has stellar mass, $M_{*} = 2.5 \times 10^{10}M_{\odot}$, a star formation rate of $14M_{\odot}$ yr$^{-1}$, and an intrisic \lya luminosity of $3.024\times 10^{44}$ erg s$^{-1}$.
The intrinsic \lya luminosity was calculated using \textsc{starburst99} (\citealt{1999ApJS..123....3L}), summing the contributions from stellar, nebular, and ISM recombination sources and assuming the following ionising and \lya radiation escape fractions: $f_{esc}^{ion}=0.02$; $f_{esc}^{Ly\alpha}=1$.
The star formation rate is comparable to the median star formation rate of the `\lya Em' sub-sample from S2011, which is $18.6 M_{\odot}$ yr$^{-1}$.

The simulations provide, as part of their output, the sites where star formation occurs during the simulation.
This is the source of the stellar feedback which drives the galactic wind.
In order to calculate the photon budget for the radiative transfer post-processing we model the star forming regions with the \textsc{starburst99} population synthesis code.
Each star-forming particle is treated as a simple stellar population (SSP), where we assume a \citet{2003PASP..115..763C} Initial Mass Function (IMF), to be consistent with the assumptions made for the hydrodynamical simulations.
We use the instantaneous star formation mode of \textsc{starburst99}, and integrate in time.
Thus for the duration of the burst we ascertain the mean ionising photon count, \lya luminosity, and the time-averaged ionising photon spectrum of the SSP.

Finally, for comparison to our fiducial model, we also run hydrodynamical simulations with the same initial conditions but with less efficient feedback.
For these runs we use the \citet{2002MNRAS.333..649S} feedback prescription which does not give an explicit kick to wind particles nor decouple them from hydrodynamics.
Throughout this paper we refer to this model as the inefficient feedback model.

\begin{figure*}
\includegraphics[width=\textwidth]{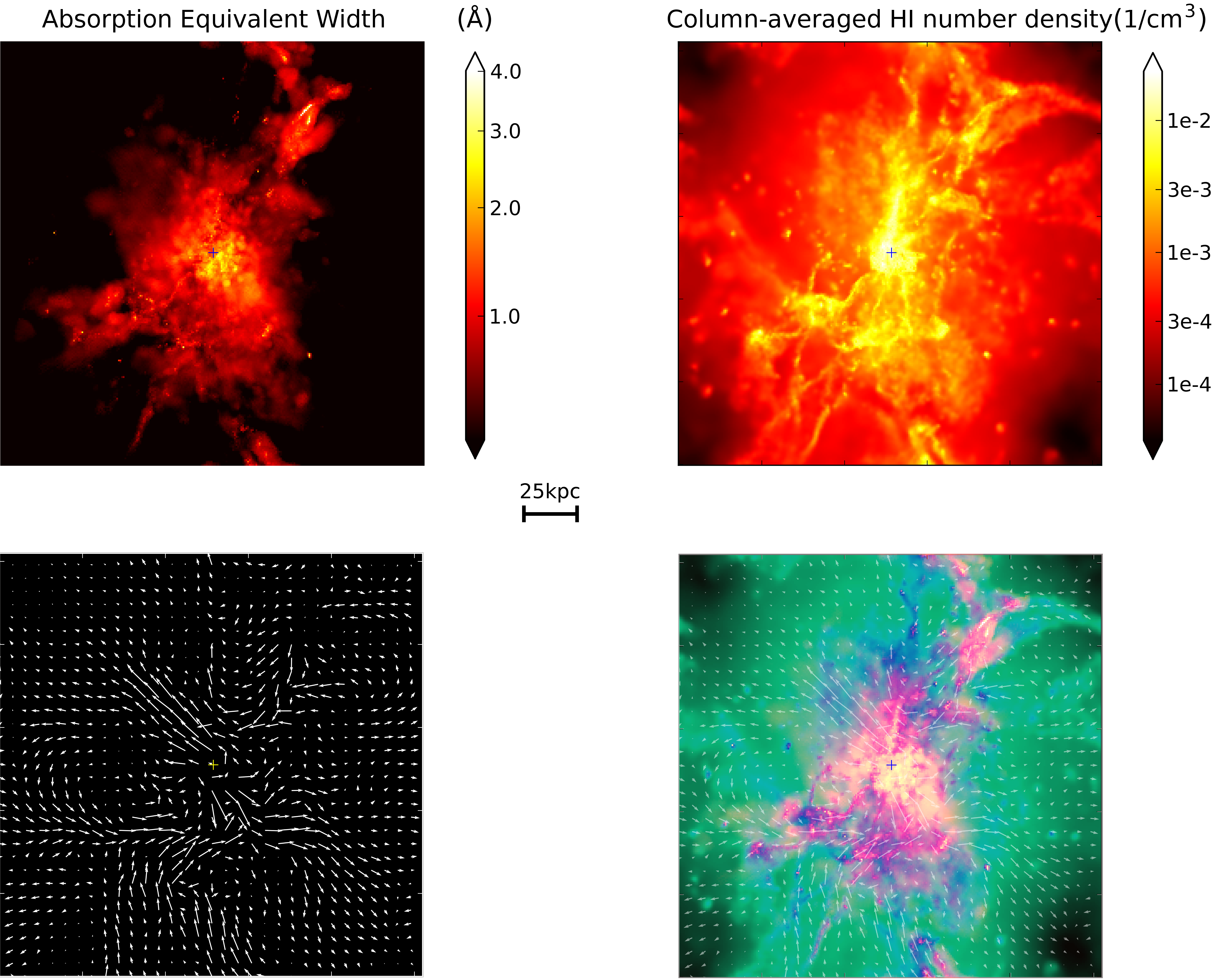}
\caption{\emph{Upper-left panel:} Synthetic equivalent width map of a $\sim$187 kpc (physical) simulation box at $z=2.2$.  The `+' symbol denotes the centre of the galaxy as defined by the gas centre-of-mass. \emph{Upper-right panel:} The same simulation box but showing column-averaged HI number density. \emph{Lower-left panel:} Depiction of the velocity field structure for the central slice (linear scale). \emph{Lower-right panel:} Composite image additively overlaying the equivalent width, density, and velocity fields of the previous 3 panels.  For clarity the column-averaged density plot has been made 30\% transparent.}
\label{fig1}
\end{figure*}

\subsection{Ionising Photons Radiative Transfer}\label{ionisingRT}
A smaller box of side $\sim 600$ comoving kpc (corresponding to $\sim$ 328 physical kpc at $z$=2.65 and $\sim$ 187 physical kpc at $z$=2.2) was cut out of the SPH simulation output and gridded onto a discrete, uniform grid with dimension $N_c=256$.
This ensures a high-enough resolution to give converged results in the radiative transfer calculations.
The output provides the temperature and density fields of the gas, as well as the location and luminosity of the stellar populations used as ionising radiation sources.
Once augmented with information on the ionisation state of the gas, it can be used as initial conditions for performing the ionising radiation radiative transfer of the galaxy's local sources.

To obtain the initial conditions for the ionisation state we assume the presence of a \citet{2012ApJ...746..125H} uniform UV background (UVB) and photoionisation equilibrium between the gas and the UVB.
With the initial conditions defined by the above procedure the effect of the local sources given by the hydro simulations is added with \textsc{crash} \citep{2001MNRAS.324..381C, 2003MNRAS.345..379M, 2009MNRAS.393..171M, 2009MNRAS.393..872P, 2013MNRAS.431..722G}, a ray-tracing Monte Carlo 3D radiative transfer code which follows the propagation of the ionising continuum and its effect on the gas it crosses.

The output of \textsc{crash} includes the temperature and ionisation state of each cell in the simulation volume.
We use $5\times10^{4}$ ionising photon packets per source and have checked that the results are converged such that the ionisation state of the simulation box is not significantly affected by using more photon packets.   
We refer the interested reader to the original papers for more details on the code \textsc{crash}.

The recombination timescale of the ionised gas is long, but nevertheless some recombination should occur.
Because the UVB is not explicity included in the radiative transfer calculation, the gas at large galactocentric radius which has been highly ionised by the initial UVB, could artificially recombine during the \textsc{crash} run.
To cope with this, we estimate which cells have a UV flux dominated by the UVB and which by local sources.
This is done by summing, for each cell, the ionising flux from all sources assuming a $r^{-2}$ falloff and comparing this total to the UVB ionising flux.
The results presented in the main body of this paper use a 1:1 domination criterion.
That is, a cell is considered to be UVB dominated if the contribution from the UVB to the the cell's ionising flux is greater than that from local sources (see \S~\ref{dom_criteria} for a discussion of this criteria).
In all subsequent post-processing steps we use either the initial ionisation state calculated under UVB photoionisation equilibrium for the UVB dominated cells, or the \textsc{crash} ionisation state for cells dominated by local sources.

\subsection{\lya Photons Radiative Transfer}
As we are neither able to resolve nor handle computationally the interstellar medium we simply remove it from the galaxy and allow the \lya photons to free-stream through the removed cells.
The effect of the ISM is then approximated by parameterizing the \lya and ionising continuum escape fractions ($f_{esc}^{Ly\alpha}$ and $f_{esc}^{ion}$ respectively).
This choice is also motivated by the main goal of the paper, i.e. an investigation of the impact of outflowing/inflowing material and not of the ISM.
We remove the ISM based on two criteria: a density threshold and a radial distance threshold.
That is, grid cells are tagged as belonging to the ISM and subsequently removed if their density is above a density threshold $n_{th}=0.5$~cm$^{-3}$ and they also lie within $\sim$10~kpc of the centre of mass of the galaxy.
This galactocentric radius threshold serves to ensure that only gas which is part of the galaxy itself is identified as ISM, and avoids removing high-density clumps in the CGM. 
In practice our results are insensitive to the exact radius threshold used.
Likewise we tested density thresholds of $n_{th}=\{0.1, 1.0\}$~cm$^{-3}$ and found very little variation in our results.
Here we assume a value of $f_{esc}^{ion}=0.02$ as in \citet{2008ApJ...672..765G}.
To perform the \lya radiative transfer we use a total of $\sim 10^5$ photon packets and assume $f_{esc}^{Ly\alpha}=1.0$, but as discussed later we renormalise the results to assume a different \lya escape fraction.

The \lya radiative transfer is performed with \textsc{Super Lyman Alpha Fighter} (\textsc{slaf}), a new code which we developed during the course of this work.
\textsc{slaf} is a Monte-Carlo \lya radiative transfer code in the vein of many previous works \citep[e.g.][]{2002ApJ...578...33Z,2005ApJ...628...61C,2006ApJ...649...14D,2006ApJ...645..792T,2006A&A...460..397V,2009ApJ...696..853L,2012MNRAS.424..884Y,2014MNRAS.444.1095G}, and can be applied to arbitrary 3D gas distributions and velocity fields.
All \lya photons are injected into the CGM at line-centre.
We have briefly investigated the effect of a Gaussian injection line profile and find that for a Gaussian with a standard deviation of 150~km~s$^{-1}$ the shape of the surface brightness profile is not altered considerably.
However, we leave the discussion of a detailed treatment of the injection profiles for a future paper.
We also developed a code, \textsc{Lyman Alpha Fighter} (\textsc{laf}), to calculate absorption line profiles for sightlines through the simulation volume, using the ionisation state and temperature output of the \textsc{crash} radiative transfer.
With these two codes we can generate emission maps (using \textsc{slaf}), and absorption equivalent-width maps (using \textsc{laf}).
Technical details of \textsc{slaf} and \textsc{laf} are discussed in Appendix \ref{appendix}.

We include a very simple dust model in our \lya radiative transfer calculations.
Dust is treated as a grey absorber with a characteristic grain radius and dust-to-hydrogen ratio tuned to satisfy the redshift-dependent dust optical depth, $\tau_{dust}$, relation from \citet{2012MNRAS.422..310G}.
This gives the dust optical depth as a function of HI column density, $\tau_{dust}(\lambda) = \left(A_{\lambda}/A_V\right)_{Z_{\odot}}  \left(Z/Z_{\odot}\right)^{1.35}  \left(N_{HI}/2.1 \times 10^{21}cm^{-2}\right) \left(1+z\right)^{-\frac{1}{2}}$.  As in \citet{2012MNRAS.422..310G}, $(A_{\lambda}/A_{V})_{Z_{\odot}}$ is the solar metallicity extinction curve from \citet{1983A&A...128..212M}, Z is the gas metallicity, $Z_{\odot}$ is the solar metallicity, and $N_{HI}$ is the HI column density.
This prescription is applied to each cell in the simulation volume with the assumption of solar metallicity everywhere.
Although this assumption is of approximate nature, our testing has shown that the results are insensitive to it.

\section{Results}\label{results}

In this section we present the results we obtain by post-processing the galaxy in the manner described above.
We break down the results by the two quantities which can be directly compared to observations.
For reference we also present the results from the inefficient feedback model, though we leave detailed comparative discussion to \S~\ref{conclusions}.

\subsection{Absorption}\label{absorption}

The upper-left panel of Figure \ref{fig1} shows a synthetic equivalent width map derived from the absorption line profiles output by \textsc{laf}.
The first step to create the map is to choose an orientation from which to view the radiative transfer simulation volume.
Then, from each cell on the far face of the box a ray is cast perpendicular to the face, towards the observer, for a total of
$N_c^2$ rays.
Each ray starts with a perfectly grey spectrum, which for simplicity we set to 1.0 such that $\{\forall \lambda: I_0(\lambda) = 1.0\}$. 
As it traverses the simulation volume the absorption line profile of the intervening HI is imprinted onto the spectrum.
This is accomplished by Doppler shifting the existing spectrum into the gas frame of each cell in turn.
For the \emph{$n^{th}$} cell the optical depth, $\tau_n(\lambda)$, is calculated for each sampled wavelength of the spectrum, and the intensity exiting the cell derived as $I_{n+1}(\lambda) = I_{n}(\lambda)e^{-\tau_n(\lambda)}$ , where $I_n(\lambda)$ is the intensity entering the cell.
Thus, when each ray reaches the observer, its spectrum is known.
From this, the equivalent width of the total imprinted absorption line for each ray can be calculated as $EW = \int_{line} [1 - \frac{I(\lambda)}{I_0(\lambda)}] d\lambda$, where $I(\lambda)$ is the final intensity reaching the observer and as noted earlier $I_0(\lambda) = 1.0$.
This results in an $N_c\times N_c$ equivalent-width map.

Our radiative transfer simulation box is 187~kpc (physical) on a side at redshift 2.2.
At this scale the IGM intervening between the edge of the box and the observer could have an impact on the calculation of $EW$, and should in principle be taken into account.
As per \citet{2011ApJ...728...52L}, this would primarily affect the spectrum blueward of line-centre.
However, since the IGM at z=2.2 is mostly ionised, we do not expect significant interaction with the \lya line.
To explicitly test the effect of the size of our radiative transfer simulation box we simulated a 2x larger box, centred on the same point, and found good agreement with the results presented here.

The equivalent width maps cannot be directly compared to observational data due to the fact that in reality not every line of sight has a sufficiently bright background galaxy or quasar along it.
In other words, the relative rarity of galaxy-galaxy pairs and galaxy-quasar pairs prohibits the creation of such a map observationally.
Nevertheless it is instructive to see what could be revealed in absorption given large numbers of background sources, especially in light of the other panels of Figure \ref{fig1}.
The upper-right panel of Figure \ref{fig1} shows the HI density averaged along the line of sight of each pixel in the map, the lower-left panel shows the projection of the velocity field of the central slice, and the lower-right panel shows a composite of all the other panels.
We can see that generally speaking the morphology of the absorption equivalent width map follows that of the underlying HI distribution.
It is clear that there is a correspondence of structures in the absorption map to the structures in the density map, such as the heavy central concentration of HI and the dense filaments.
Examination of the velocity field (lower panels) shows material inflowing along the filaments, and being blown out in a biconical outflow.
Although the absorption map and velocity fields are somewhat messy, the lower-right panel shows how the velocity field (and hence outflow) affects the absorption map.
Unsurprisingly, since the absorption depends on both the density and velocity field, larger velocities tend to coincide with larger absorption equivalent widths.
Figure \ref{fig1}, in particular the lower-right panel, demonstrates the complex interplay between these three quantities.

\begin{figure}
\includegraphics[width=84mm]{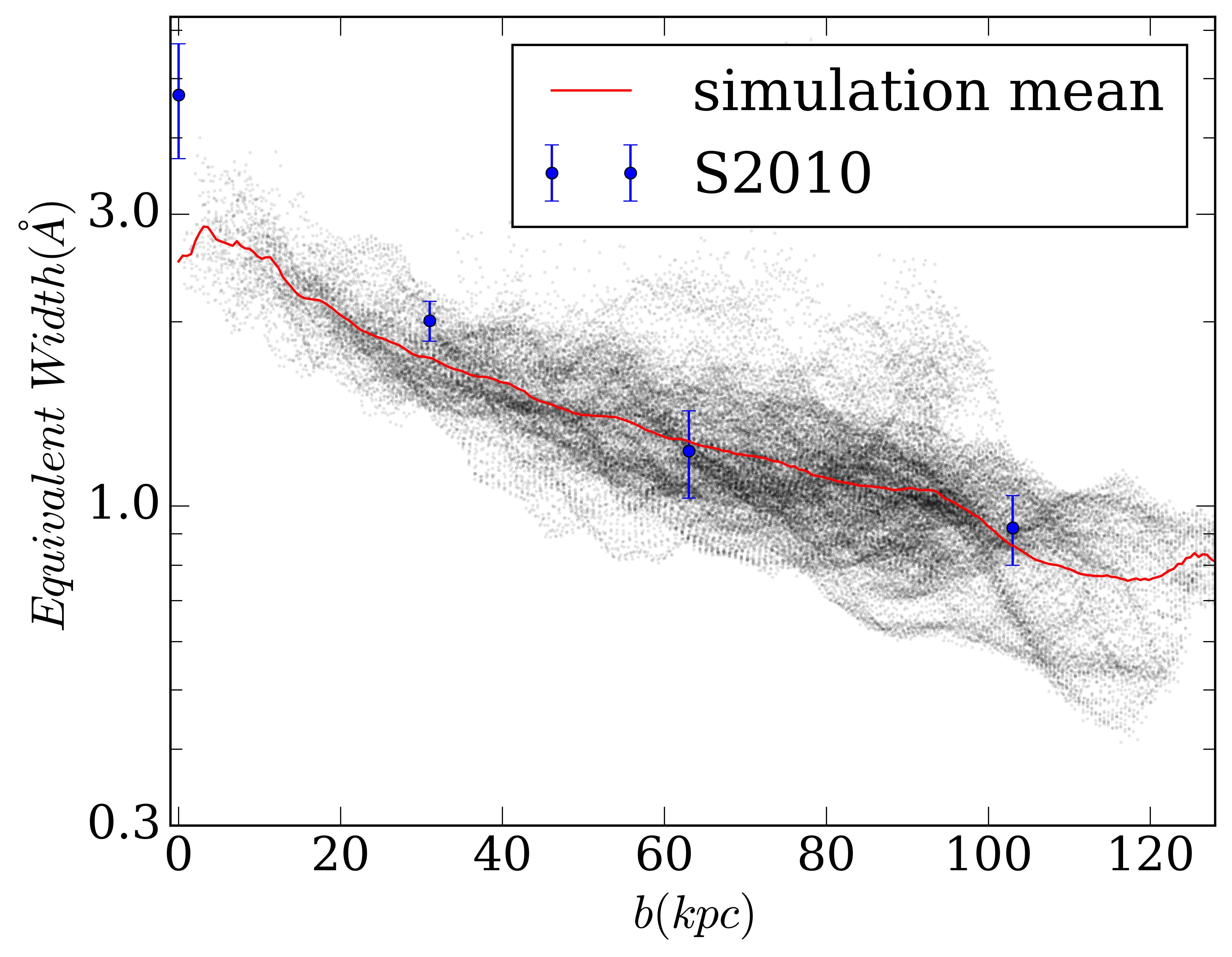}
\caption{Equivalent width ($EW$) vs. impact parameter ($b$) profile derived from the upper-left panel of Figure \ref{fig1}.
Each pixel is plotted in the $EW$-$b$ plane as a black point, with the mean profile shown by the solid red line.
Observations from S2010 are shown in blue.}
\label{fig2}
\end{figure}

Figure \ref{fig2} is a plot of absorption equivalent width vs impact parameter, $b$, i.e. an absorption equivalent width radial profile generated from the absorption map in Figure \ref{fig1} (upper-left).
It was created by sampling points in the map radially from the centre of the galaxy (denoted by a `+' in Figure \ref{fig1}).
Each pixel from the absorption map in Figure \ref{fig1} is plotted in Figure \ref{fig2} as a black point.
The solid red line shows the equivalent width profile obtained by azimuthal averaging of the map.
The blue points and error bars are the data points from the galaxy-galaxy pairs in S2010.
Figure \ref{fig2} shows that for this viewing angle our simulation provides good agreement with the S2010 observations from $\sim40$~kpc outwards, mildly underpredicts absorption compared to the data point at $b=31$~kpc, and increasingly underpredicts the absorption towards the centre of the galaxy.

\begin{figure}
\includegraphics[width=84mm]{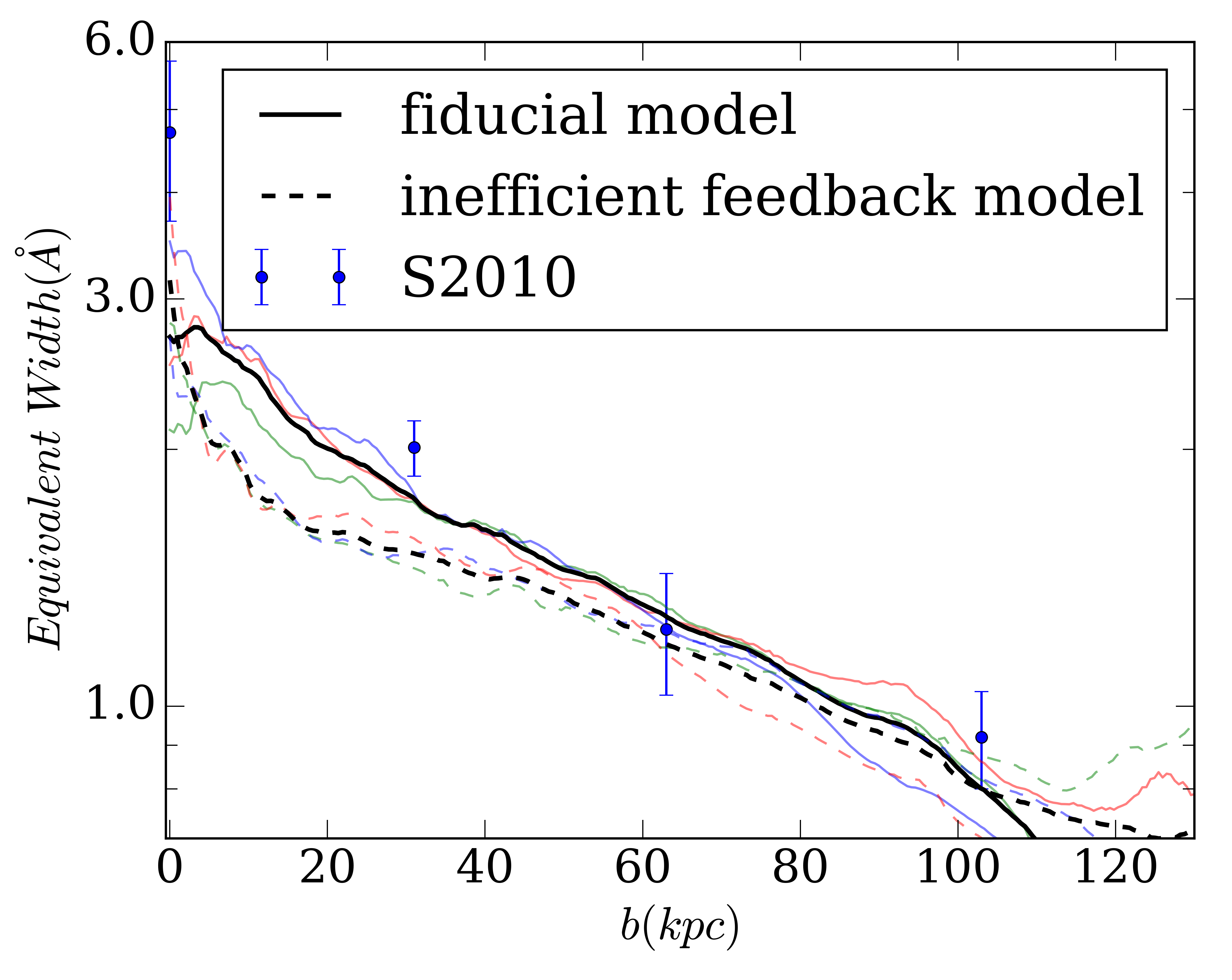}
\caption{Comparison of equivalent width vs. impact parameter ($b$) profiles from simulations with (solid lines) and without (dashed lines) strong stellar feedback.  Green, red, and blue lines are profiles as viewed along the x, y, and z axes respectively.  Black lines are the mean profiles of the three primary axes. 
Observations from S2010 are shown as blue points and error bars.}
\label{fig3}
\end{figure}

Figure \ref{fig3} shows the absorption equivalent width profile as viewed from all three axis-aligned orientations as green, red, and blue solid lines.
The red line in Figure \ref{fig3} corresponds to the data in Figure \ref{fig2}.
The mean of the three orientations is shown as a black solid line.
We note that the mean equivalent width profile from our simulations is in quite good agreement with the observations.
Similarly to the single orientation presented in Figure \ref{fig2}, the simulation is within the observational error bars for the outer 2 observational points, $\sim12\%$ below at $b\sim 31$~kpc and $\sim43\%$ below at $b\sim 0$.
The lower absorption equivalent width predicted by our simulations in the inner $\sim40$~kpc of the galaxy is indicative of the simulated HI gas density being too low, the simulated velocity dispersion being too low, or a combination of both.
We return to discuss this topic later in \S~\ref{conclusions}, taking into account the emission results presented in \S~\ref{emission}.

For comparison, the inefficient feedback model is also shown in Figure \ref{fig3} with dashed lines.
Since this model lacks strong stellar feedback the galaxy has formed more stars and thus has a stellar ionising photon luminosity around 4 times that of the galaxy formed with the fiducial model.
In order to make a meaningful comparison then, we rescale the luminosity to match that of the galaxy formed under the strong feedback model.

Figure \ref{fig3} shows that the inefficient feedback model also does a good job of reproducing the observations in the outer CGM.
However it shows markedly worse agreement with observations in the inner $\sim40$~kpc, consistently underpredicting the absorption equivalent width.

\subsection{Emission}\label{emission}\
Summing the luminosity of the photon packets exiting the simulation box, with the assumptions that the \lya escape fraction is 100\% and that all \lya radiation is emitted by the stars in the simulation, gives an observed \lya luminosity of the simulated galaxy of $3.023\times 10^{44}$ erg s$^{-1}$.
This is extremely close to the intrinsic luminosity (see \S~\ref{simulations}).
In the simulations of the \lya emission of our galaxy, we ignore the component of the signal arising from recombination radiation in the CGM.
We justify this by noting that a calculation of the recombination rate, and the resulting \lya emission, shows that the contribution from recombination within the simulated volume is less than 10\% of the total \lya emission\footnote{see \S~\ref{appendix_recombination} for further discussion}.

\begin{figure}
\includegraphics[width=84mm]{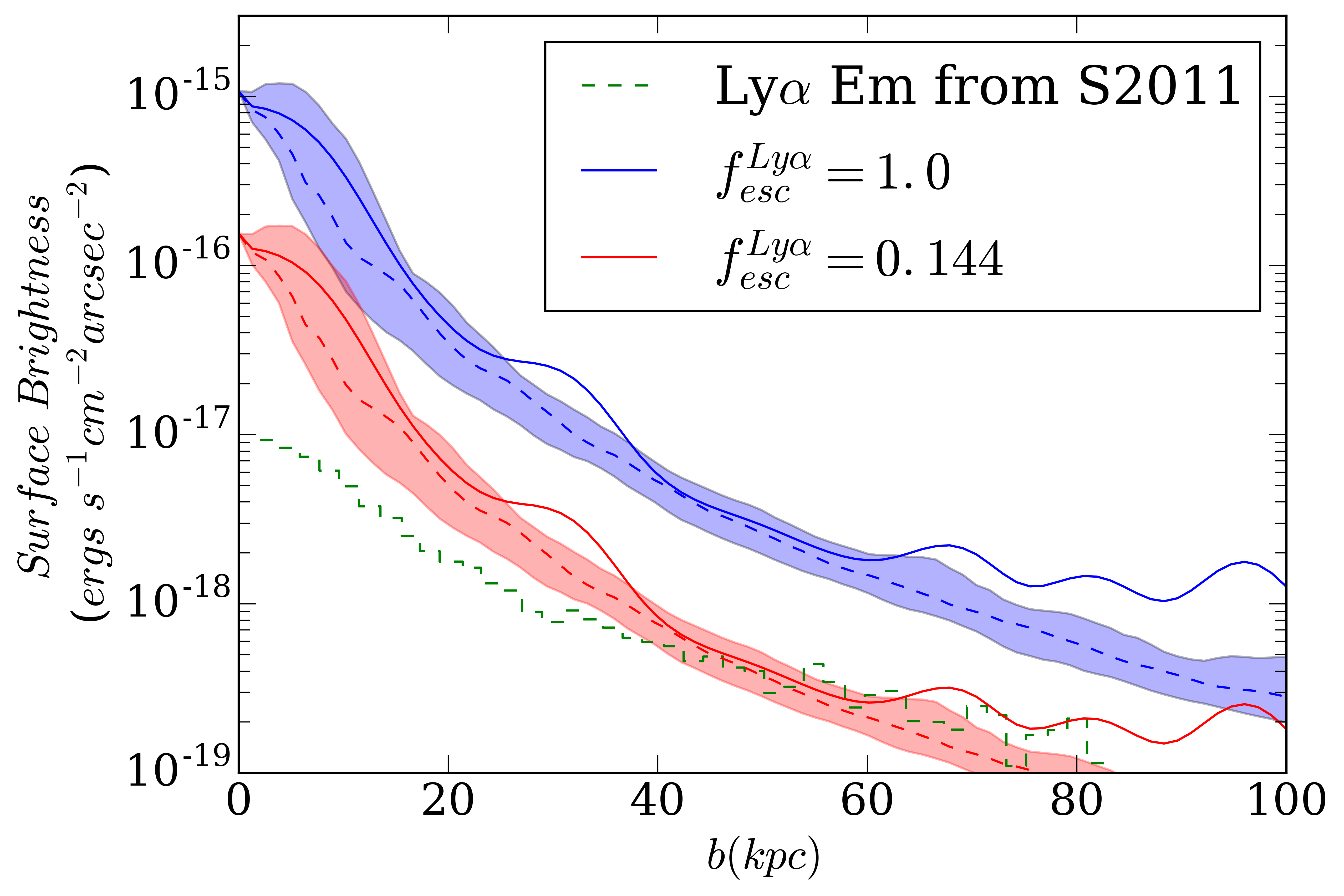}
\caption{\lya surface brightness as a function of impact parameter for our simulated galaxy at $z=2.65$.
The blue lines are for $f_{esc}^{Ly\alpha} = 1.0$, while the red lines assume $f_{esc}^{Ly\alpha} = 0.144$.
In both cases the solid line is the azimuthal mean, the dashed line is the azimuthal median, and the shaded regions show the interquartile range.
The dashed green line shows the `\lya Em' sub-sample from S2011.}
\label{fig4}
\end{figure}

\begin{figure}
\includegraphics[width=84mm]{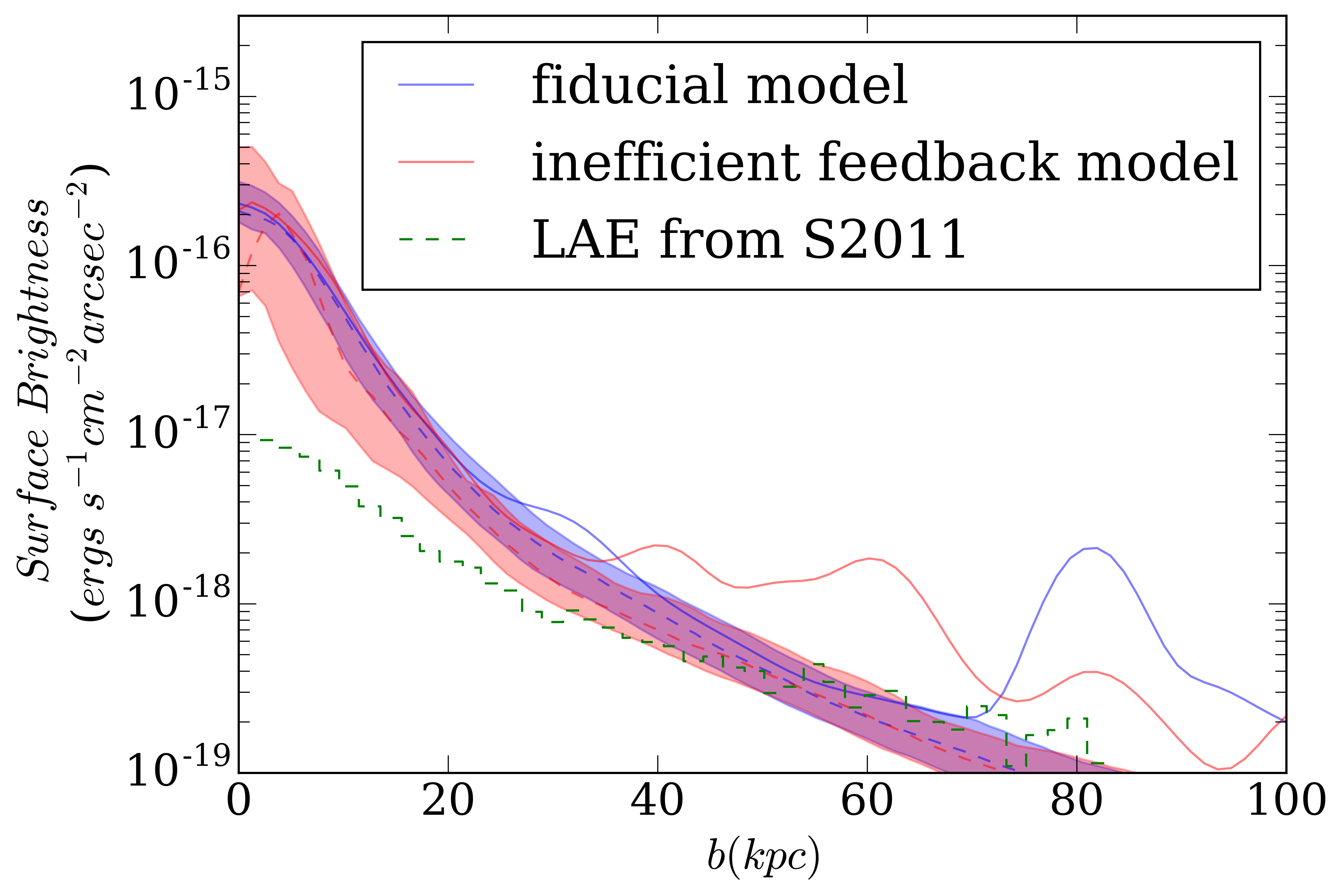}
\caption{Six orientations of the simulated galaxy stacked to give the average surface brightness as a function of impact parameter.
The blue lines show the results for our wind model whereas the red lines show the results of the galaxy simulated without strong feedback.
In both cases $f_{esc}^{Ly\alpha} = 0.14$ as per S2011.
In both cases the solid line is the azimuthal mean, the dashed line is the azimuthal median, and the shaded regions show the interquartile range.
The dashed green line refers to the `\lya Em' sub-sample from S2011.}
\label{SBprofileStack}
\end{figure}

Figure \ref{fig4} shows the \lya surface brightness profile of the galaxy as viewed from a single side.
The data in red assumes $f_{esc}^{Ly\alpha}=0.144$, i.e. 14\% of the \lya photons generated in the galaxy diffuse through the ISM and escape into the CGM without attenuation.
This is motivated by S2011, which quotes the \lya escape fraction as 14.4\% for the galaxy sub-sample we compare to.
The data in blue show the upper limit of $f_{esc}^{Ly\alpha}=1.0$.
Changing $f_{esc}^{Ly\alpha}$ simply modulates the total energy of the \lya photons injected into the CGM.
Since we use enough photon packets to sample the radiative transfer, we similarly modulate the energy assigned to each photon packet to vary $f_{esc}^{Ly\alpha}$, which has the result of shifting the normalisation of the points in Figure \ref{fig4}.

Here, solid lines represent the mean radial profile (which is the observational quantity plotted in S2011) of the simulated surface brightness images, dashed lines represent median radial profiles, and the shaded regions show the lower and upper quartiles.
The simulated surface brightness images are degraded to 1'' FWHM resolution prior to creating these profiles, in order to match the resolution of the S2011 data, represented by the green dashed line.

Interestingly, at times the mean profile rises above the median profile, and even above the upper quartile of the distribution at some impact parameters.
This is indicative of the fact that at these impact parameters the distribution is not Gaussian but skewed, dragging the mean up.
This is caused by a small amount of substructure in the surface brightness images - small, bright star-forming clumps.
The impact parameter at which these clumps reside depends on projection effects, and thus on the observation angle.
In this paper we only simulate one galaxy but given the random projection of substructure we expect that a fairer comparison, stacking many simulated galaxies, would have a similar mean to the median.
This is expected because the effect of outlying substructure at a given impact parameter would be diluted when averaged over more galaxies, most of which would not exhibit substructure at the same impact parameter.

We argue then that in the case of Figure \ref{fig4} where there is little substructure it is perhaps better to compare the median profile to the results from S2011, which are themselves a stack of 52 galaxies.
Considering the median profile in Figure \ref{fig4} it is clear that our simulation is a good fit to observations from $\sim40$~kpc out to where the observations end at 80~kpc.
Below $\sim40$~kpc the surface brightness profile given by our simulation starts to rise above the observed profile and becomes too peaked.
The model proposed in \citet{2012MNRAS.424.1672D} also exhibits a similar behaviour in the inner region.

Figure \ref{SBprofileStack} shows the average surface brightness profile obtained by stacking the six surface brightness maps corresponding to viewing the simulated galaxy from the six sides of the simulation volume.
When stacking several viewing angles the shape of the profile remains very similar as there is suprisingly little variation in the profile when the galaxy is viewed from different orientations.
The main difference between the profiles seen from different orientations is in the location and size of the bumps and peaks.
The large bump at 80~kpc comes from a bright peak visible along a single axis.

Figure \ref{SBprofileStack} also shows the inefficient feedback model in red.
As was done for the absorption results, the stellar luminosity of the inefficient feedback model has been scaled down to match that of the fiducial model.
The medians of the two models are almost indistinguishable, thus the median surface brightness profile from the inefficient feedback model is also compatible with the data from $\sim35$~kpc to 80~kpc.
However, whereas the fiducial model has good agreement between the mean and median for the vast majority of the profile, the mean of the inefficient feedback model is significantly above the median from $\sim35$~kpc outwards. This discrepancy arises because in the inefficient feedback model there is much more substructure over a large range of impact parameters compared to the fiducial model, in which there are only a few, isolated clumps.
From this single realisation of a galaxy it is not possible to tell if the inefficient feedback model generically predicts an increased presence of substructure, but if this were indeed the case, then we argue that here the median is not a good indicator of what can be expected from mean-stacking many simulated galaxies.

\begin{figure}
\includegraphics[width=84mm]{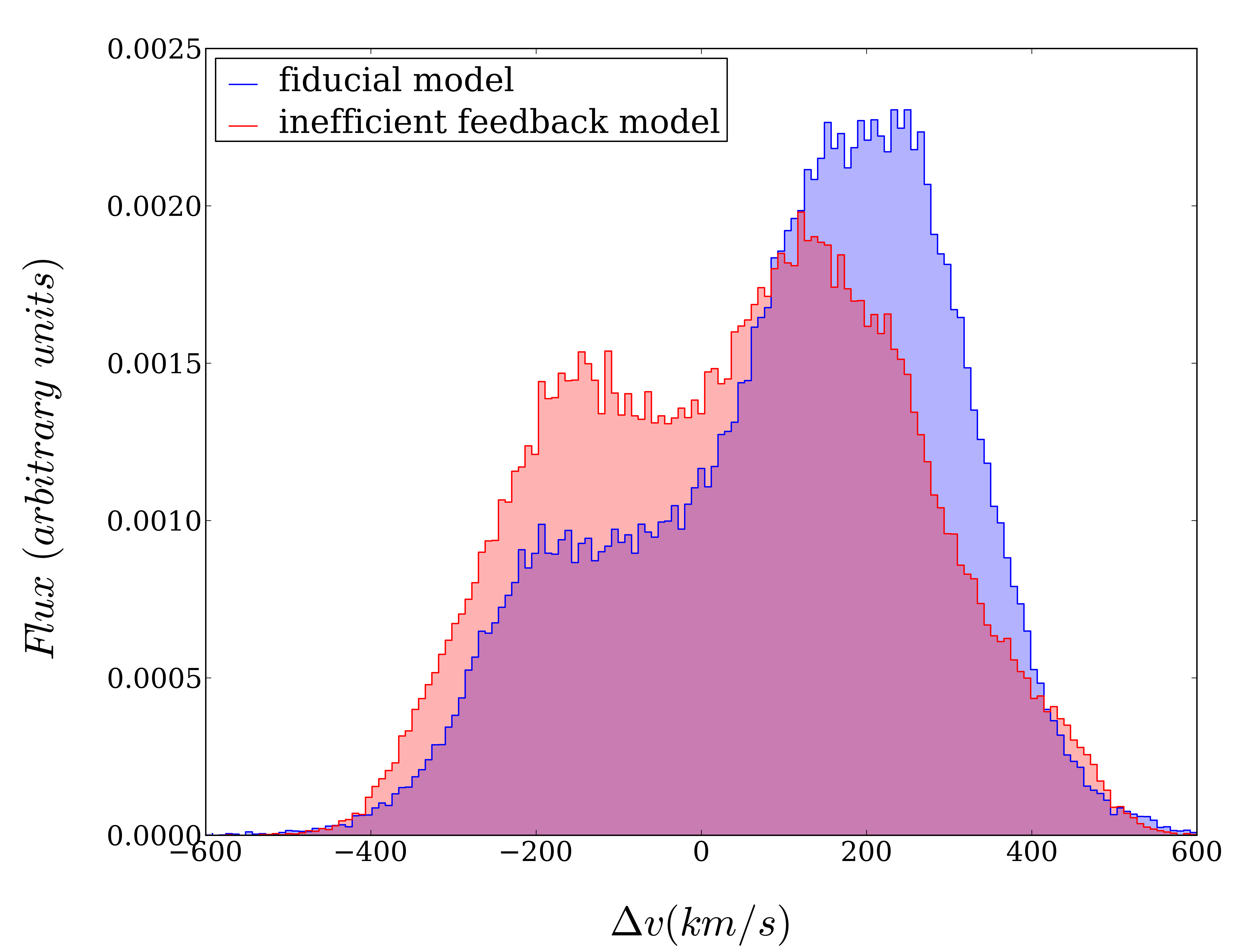}
\caption{Spectrum of all \lya photons exiting the simulation box for the fiducial (blue) and the inefficient feedback (red) models.
The y-axis is in arbitrary units, and each spectrum has been normalised to unity.
The x-axis is the velocity offset from the systemic velocity.
In this spectrum the sign convention is such that a positive velocity offset corresponds to a negative frequency offset.
}
\label{spectrum}
\end{figure}

Figure \ref{spectrum} shows the spectrum of all \lya photons exiting the radiative transfer simulation box.
There is a clear distinction between the two models.
The fiducial model has a more pronounced redward peak, and a reduced blueward peak relative to the inefficient feedback model.
This can be understood as a less extreme example of the effect shown in Figure \ref{slafDynamicSpheresTest}, whereby an expanding medium boosts the redward peak and suppresses the blueward peak.
Since the fiducial model has stronger feedback, and thus a stronger outflow, this is exactly the behaviour we expect to see in the spectrum. The larger velocity shift of the red peak has the same origin. 
As we discuss in \S~ \ref{conclusions} the velocity offset of the peaks in both models is too low compared to observations, but aside from that, the line shape of the fiducial model is generally in better agreement with observations (\citealt{2010ApJ...717..289S}).

\subsection{Combined Absorption and Emission}\label{combined}
Comparing our fiducial model with the inefficient feedback model we can state that judged by absorption alone the former better fits observations, while if we consider emission only the message is less clear.
Assuming that for both models the median surface brightness profile is a good proxy for what would result from mean-stacking many more simulated galaxies, then neither model is a better fit to observations than the other.
However, this assumption may not be true, and instead the mean profile of the inefficient feedback model could be a better proxy for the mean-stacked profile.
If this is the case then the fiducial model again provides a better fit to observations.

Taking both absorption and emission into account it is clear that the fiducial model is favoured, with the caveat that this study is limited to a single simulated galaxy.
At this stage, this preference comes primarily from the absorption result.
As explained above, whether the emission actively supports this preference or merely does not contradict it hinges on the results of simulating more galaxies.

\section{Discussion}\label{conclusions}
We have demonstrated that our fiducial wind model can reproduce some of the observables detailed in S2010 and S2011, although it struggles in the inner regions.
At first glance, the partial success of the outflow model is surprising since it is by necessity extremely simplified.
In particular, it has the non-physical characteristic of temporary decoupling of the kicked wind particles, which is necessary to produce efficient feedback.
In fact, \citet{2008MNRAS.387.1431D} suggest that the density threshold at which hydrodynamical coupling of the kicked particles to their neighbours is reinstated may be the dominant factor in determining the morphology of this class of outflow models.

In more detail, in \S~\ref{absorption} we noted that the simulated absorption equivalent widths within the inner $\sim40$~kpc of the galaxy are lower than observations, suggesting that in this region either the simulated HI gas density is too low, the simulated velocity dispersion is too low, or a combination of both.
In \S~\ref{emission} we saw that the simulated emission profile is too peaked compared to observations, i.e. the \lya photons are arriving at the observer too directly (to see this intuitively consider the case of a point source in a vacuum - with no scattering the observer would see a 1-pixel wide step function emission profile).
The emission surface brightness profile results confirm one of the options from the absorption result: a higher gas density would lead to more \lya photon scatterings, flattening the emission profile as required to match observations; a higher velocity dispersion would Doppler shift the HI away from \lya resonance, allowing the \lya photons a more direct path to the observer and further increasing the peakiness of the emission profile.
Thus the emission surface brightness profile results suggest that a more successful wind model should have a higher central HI gas density.
However, this alone cannot be the whole explanation.
In both cases the emission spectra (Figure \ref{spectrum}) show a red peak at a velocity offset less than that observed ($\sim$400~km/s, see \citealt{2010ApJ...717..289S}).
This is indicative of the outflow velocity not being high enough.
It appears then that a combination of both increased HI density in the inner regions and increased outflow velocities is needed to bring simulations into line with all of the available observations.

While we need further simulations to confirm this, we speculate that since both of the tested models show similar absorption and emission profiles at $b>50$~kpc, the CGM signal in \lya absorption and emission may be largely unaffected by outflows at these impact parameters.
Indeed, \citet{2013ApJ...765...89S} show that outflows do not disrupt cold inflows.
Thus, if the signal at $b>50$~kpc is dominated by cold accretion flows it is reasonable to expect that our strong feedback model would not significantly affect the absorption and emission profiles in this region.
However, we caution against interpretating the fact that the $b>50$~kpc signal does not appear to be coming from outflows to mean that it must necessarily be caused by inflows.

This limited radial influence of the outflow is perhaps a manifestation of the fact that a wind has some finite sphere of influence.
Visually, we can see this by referring to the lower-left panel of Figure \ref{fig1}, where we can see that beyond $\sim50$~kpc the amplitude of the velocity field drops off.
The extent of this sphere will surely depend on the velocity of the stochastic kick given to wind particles, but may also be affected by the aforementioned density threshold at which wind particles become re-coupled to hydrodynamics.

So far we only present the results of one simulated galaxy, and it would be premature to judge the fiducial outflow model on this one result.
Since S2010 and S2011 both deal with averaged/stacked data the correct comparison to make is to similarly stacked simulated galaxies.
\citet{2012ApJ...745...11G} simulate a suite of galaxies and we suggest a stacked analysis of this data would be a significant improvement on the work presented in this paper.

In fact, the need for more simulations is deeper than may be immediately obvious.
The two different feedback models we have presented produce galaxies with different properties from the same halo initial conditions.
This, of course, is exactly why we are interested in various feedback models, but as we saw in \S~ \ref{results}, it makes comparison of the resulting galaxies difficult.
Recall that the `re-simulated' halo was chosen such that the galaxy which forms under the fiducial feedback model was similar to the mean of the S2010/S2011 samples.
Using the inefficient feedback model the galaxy which formed was brighter and so we needed to rescale the luminosity of the galaxy to compare to the fiducial model and observations.
This introduces an inconsistency: the stars which ionise/illuminate the CGM in \lya do not have the same properties as those which generated the outflow velocity field and CGM properties.
If we instead chose to compare two galaxies with the two models, based on the final formed galaxy properties (that is, choose galaxies from both models to match observations) we are faced with the problem that we would be comparing galaxies with different initial conditions.
Since what we are really interested in is the impact of the different models on the evolution of the CGM we prefer the approach we have taken where we keep the initial conditions the same across the two models.
Nevertheless, we acknowledge that this is not ideal and suggest that the real solution is to simulate a large statistical sample of galaxies for both models, from which a sample with properties matched to observations can be selected and compared.

Other outflow models such as those described in \citet{2012MNRAS.426..140D} and \citet{2008MNRAS.387.1431D}, avoid the unphysical temporary decoupling from hydrodynamics common to \citet{2003MNRAS.339..289S}-style winds.
The resulting outflows are qualitatively different in spatial and velocity morphology from those produced by \citet{2003MNRAS.339..289S} and would thus make for an interesting comparison.
Additionally, there are free parameters in the outflow model used by \citet{2012ApJ...745...11G} which presents a parameter space which it may prove fruitful to explore.
A comparative test of outflow models can be found in \citet{2016ChungThesis}

Our simulations do not include spatially extended emission from cold accretion streams, either powered by recombination or collisional excitation. Previous works have shown that this may provide a large \lya luminosity \citep[e.g.][]{2009MNRAS.400.1109D}.

Accounting for this emission may boost the amount of \lya emission at large impact parameters, thus flattening the predicted surface brightness profile.

Our simulations show that a stronger outflow appears to diminish the `twin peaks' profile of the \lya line, emphasising the red peak and diminishing the blue peak.
Recent observations by \citet{2017A&A...608A...8L} with MUSE\footnote{\url{http://www.eso.org/sci/facilities/develop/instruments/muse.html}} include spectra with similar asymmetric peaks, which may be attributable to the processes discussed in this paper.

This work has possible relevance to Enormous \lya Nebualae \citep[][]{2017ApJ...837...71C} and the enigmatic \lya blobs reported by \citet{2000ApJ...532..170S} and \citet{2004AJ....128..569M}.
It is presently unclear what the physical mechanism powering these objects is.
One possibility is that \lya blobs are powered by a central galaxy with the photons scattered off the surrounding medium, in a similar fashion to that considered in this paper.
\citet{2011Natur.476..304H} presents observations supporting this model of \lya blobs, but see also \citet{2000ApJ...537L...5H}

\subsection{Comparison with Previous Work}\label{comparison}
As we have already pointed out, the absorption line data is better fit by our fiducial model, but the inefficient feedback model also does a good job of reproducing the observations at higher impact parameters.
The similarity of the profiles in Figure \ref{fig3} goes some way towards explaining why previous analyses of \citet{2012MNRAS.424.2292G} and \citet{2013ApJ...765...89S} can reproduce the absorption line data without strong outflows, instead attributing most of the \lya absorption to inflowing cold streams.

\citet{2012MNRAS.424.2292G} reproduces the \lya absorption line data extremely well, even towards the centre of the galaxy.
This is in contrast to our inefficient feedback model, which underpredicts the amount of \lya absorption in the inner $\sim40$~kpc.
One reason for this may be the fact that \citet{2012MNRAS.424.2292G} do not include the effect of local sources, and instead use a simple self-shielding criteria to calculate the ionisation state of the gas assuming a UV background.
It is therefore reasonable that our simulations which contain local sources, mostly concentrated in the galaxy, and explicitly compute the ionisation state, should have a higher ionisation fraction in the central region close to the galaxy.
This would naturally lead to a lower \lya absorption equivalent width.
In the outer regions, where local sources are less dominant, we expect better agreement with the \citet{2012MNRAS.424.2292G} result and indeed this is exactly what we see.

An alternate explanation for the discrepancy could be the way in which we calculate the initial conditions for our \textsc{crash} runs.
Whereas we assume all gas is in photoionisation equilibrium with the UVB (with no shielding), \citet{2012MNRAS.424.2292G} use a self-shielding criteria. 
In the dense regions towards the centre of the galaxy this may cause us to overestimate the ionisation fraction of the gas.
If this is happening, then it is also happening in our fiducial model, which also lies under the observations at very small impact parameter, and a better treatment of the UVB in our calculations may yield better agreement with observations for our strong feedback model.

\citet{2013ApJ...765...89S} does include local sources, albeit in a simplified fashion, placing all sources at the centre of the galaxy.
They have problems similar to ours in reproducing the central two absorption data points from S2010.

With respect to emission, S2011 presents a simple analytic model to explain their observations.
They consider a spherically symmetric outflowing HI CGM, modulated by a covering fraction which is a power law function of galactocentric radius.
Radiative transfer is treated with an extremely basic prescription.
Nevertheless this model provides a good fit to their observations.
The success of this model provided motivation for us to test whether the underlying assumption of a central source, emitting radiation which scatters in an outflowing CGM, could stand up to the scrutiny applied by our hydrodynamic simulations and a full treatment of radiative transfer.

\citet{2009ApJ...696..853L} perform a similar radiative transfer treatment to this work.
However, their work differs in some important ways.
The galaxies they simulate are not targeted to match the sample of galaxies from S2011,
as they are at a different redshift ($z$=3.6) and with no strong feedback.
Nevertheless, their results are comparable with a very similar surface brightness profile shape. 

Finally, on the observation side, it would be disingenuous to omit mention of recent work by \citet{2015Sci...348..779H} and \citet{2014Natur.506...63C} which cast doubt on the fidelity of current numerical cosmological simulations of the type used in this work.
\citet{2015Sci...348..779H} use background quasar absorption lines to study the properties of foreground quasar hosts.
They observe much more cool gas than simulations predict, suggesting that `essential aspects' of massive halo hydrodynamics at $z \sim 2$ are not being captured by current cosmological simulations.
\citet{2014Natur.506...63C} use fluorescent \lya emission to study the cold gas mass of the nebula surrounding quasar UM287, and again find an excess of cold gas relative to simulations.
While the case against cosmological simulations brought by these works gives cause for concern, it is consistent with our findings as discussed above - namely that additional HI is required to bring our simulations into agreement with the observations. 

\section{Summary}
We have introduced a new test for galactic outflow models, which combines hydrodynamical simulations and \lya radiative transfer in a self-consistent way where the stars driving the outflows are also responsible for the ionising and \lya radiation used in the radiative transfer.
Crucially, we use constraints from both \lya absorption and emission to test our models.
The fiducial outflow model which we have presented in this paper can reproduce features of both absorption and emission observations, although the inner region and the \lya emission spectrum remain problematic.

We also showed that there are differences in these two diagnostics when a different feedback model is used.
Furthermore, comparison of our results for the two feedback models hints that galactic outflows may predominantly affect the inner $\sim$50~kpc of the CGM.
This suggests that future \lya observations of the inner CGM may be key to gaining a better understanding of the galactic outflows which appear to be important to galaxy formation and evolution.

\section*{Acknowledgements}
We thank Luca Graziani for countless fruitful discussions during this work.
Many thanks also to Akila Jeeson-Daniel for providing the \textsc{crash}-$\alpha$ pipeline, and all the helpful support along the way.
We also thank Kathryn Kreckel, Tyrone Woods, Andressa Jendreieck for constructive comments.
Finally, we thank the reviewer Joakim Rosdahl for insightful comments which, when addressed, surely improved the quality and robustness of this paper.

\appendix
\section{code details}\label{appendix}
In order to perform this work it was necessary to develop some new codes.
Here we briefly describe the nature of the two codes, and show the results of some of the verification tests we have applied to them.

\subsection{Radiative Transfer (SLAF)}
\textsc{slaf} (Super Lyman Alpha Fighter) is a new 3D monte-carlo \lya radiative transfer code similar to \citet[][]{2002ApJ...578...33Z,2005ApJ...628...61C,2006ApJ...649...14D,2006ApJ...645..792T,2006A&A...460..397V,2009ApJ...696..853L,2012MNRAS.424..884Y}.
In order to strike a balance between efficiency, extensibility, and maintainability, C++ was chosen as the implementation language, and an object-oriented design paradigm was used. 

The papers cited above describe in detail how codes of this nature work.
However, here we give some information specific to our code and briefly outline the critical steps involved in \lya RT.
\textsc{slaf} was designed to fit into the existing \textsc{crash} pipeline.
Therefore, the input files share the same format as \textsc{crash}.

A number of photon packets are allocated to each source, based on the source luminosity.
Each photon packet is then tracked from the source until it exits the simulation box.
The first step is to choose a random direction in which to emit the packet from the source.
To propagate the photon packet from scattering to scattering we need to randomly choose an optical depth, $\tau$, that the photon will freely stream through from the probability distribution $e^{-\tau}$.
We then propagate the photon packet from cell to cell, calculating the optical depth `used up' in traversing each cell, based on the frequency of the photon and the physical properties of the cell.
As the photon packet freely streams from cell to cell, a dust attenuation factor is applied, based on the HI density of the cell and a specified dust-to-gas ratio.
When the optical depth of the photon packet is exhausted the photon undergoes a scattering event.

For each scattering event we must first generate the thermal motion of the atom which we are scattering off.
This is done by picking the thermal velocity from a non-trivial distribution, details of which are given in \citet{2002ApJ...578...33Z}.
In the current version of \textsc{slaf} scattering is isotropic.
The direction of the scattering together with the thermal motion of the atom determines the gas-frame frequency shift of the scattered photon packet.
We are then ready to begin the loop again and propagate to the next scattering event.

To generate surface brightness images we follow the method of \citet{1984ApJ...278..186Y}.
We set up virtual detectors on each face of the simulation box.
Then, for each scattering event we calculate the probability that the photon scatters in the direction of each `detector'.
In our case this means that we calculate the probability of scattering in the direction of the 6 faces of the simulation box.
For each scattering event and for each detector the energy of the photon at the time of the scattering event is weighted by the probability of scattering towards said detector and added to the pixel that the scattering event projects onto at that detector.
At the end of the radiative transfer each detector has a relative brightness map.
In order to convert these relative brightness maps to absolute values we sum the total energy of all photons which exit the simulation cube and assign it proportionally over the maps.

\begin{figure*}
\includegraphics[width=\textwidth]{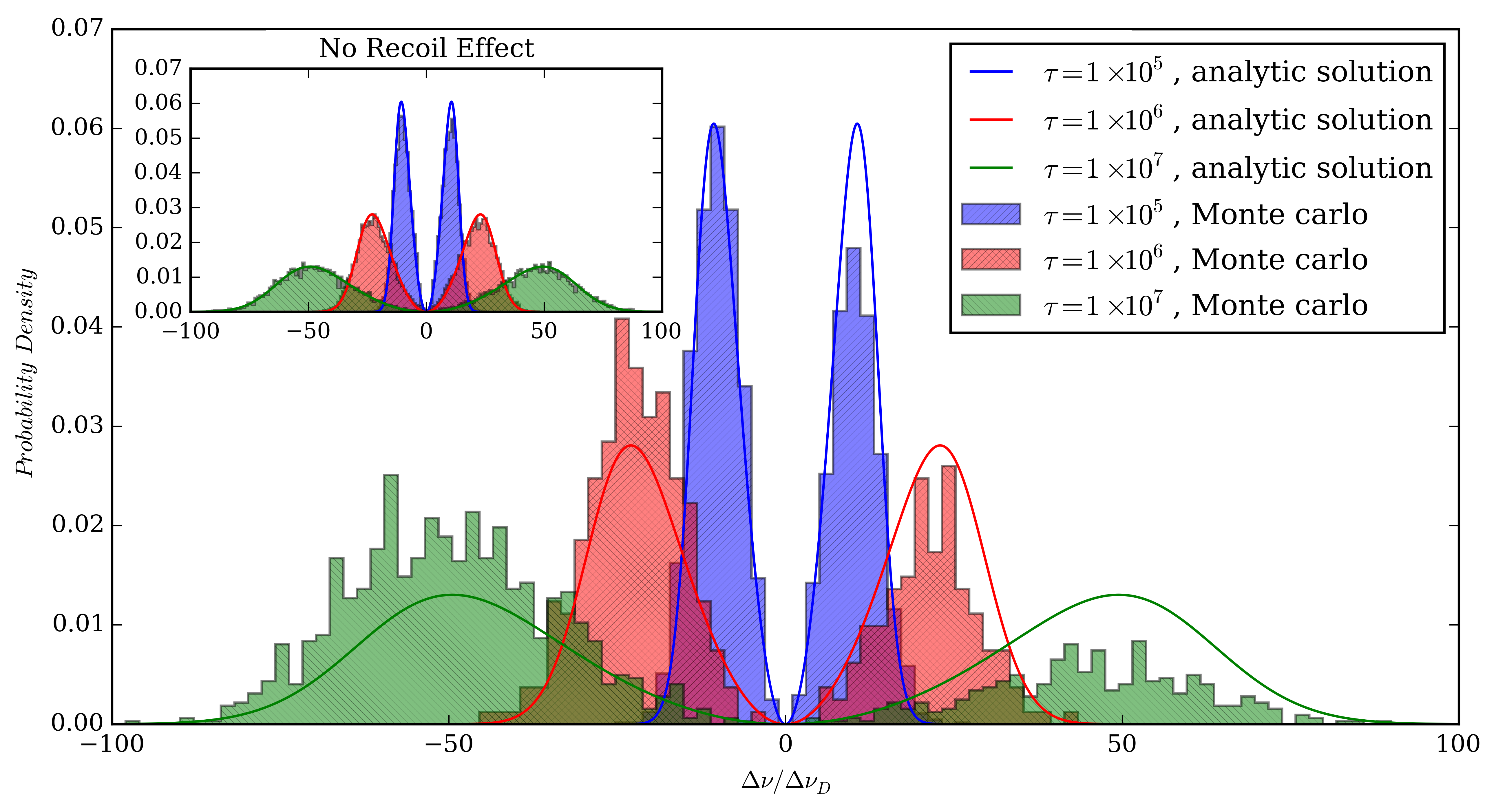}
\caption{Comparison of the spectrum calculated by \textsc{slaf} (histograms) with a known analytic approximation (lines) for a line-centre optical depth of $10^5$ (blue), $10^6$ (red) and $10^7$ (green).
The apparent discrepancy is due to the omission of the `recoil effect' in the analytic approximation.  
$\Delta\nu\equiv\nu - \nu_0$ is the frequency shift; $\Delta\nu_D \equiv (\frac{v_{th}}{c})v_{0}$ is the Doppler frequency shift, where $v_{th}=(\frac{2k_bT}{m_H})^{\frac{1}{2}}$}; $k_B$ is the Boltzmann constant; $m_H$ is the mass of the Hydrogen atom; $T$ is the temperature.
The inset plot shows the output of our code with the `recoil effect' disabled. This simulation run was performed with T=10K on a 128$^3$ grid.
\label{slafStaticSphereTest}
\end{figure*}

In order to verify our implementation, we compare the \textsc{slaf} output with the test case in \citet{2006ApJ...649...14D}, for which there is an approximate analytic solution \citep[][Equation 9]{2006ApJ...649...14D}.
In brief this test puts a single, monochromatic \lya source at the centre of a homogeneous static sphere of HI gas at 10~K.
In each instance the uniform gas density is chosen so that the line-centre optical depth from the source to the edge of the gas sphere is \{$10^5$, $10^6$, $10^7$\}.

The outcoming spectrum from \textsc{slaf} is compared to the analytic solution in Figure \ref{slafStaticSphereTest}.
\textsc{slaf} is shown to be in excellent agreement with the analytic solution with the exception of a systematic amplification of the red peak and suppression of the blue peak compared to the analytic solution.
We understand this to be due to the `recoil effect' \citep[][]{1959ApJ...129..551F, 2002ApJ...578...33Z, 2006ApJ...645..792T}, the thermalisation of photons which occurs due to the photon scattering in a different direction to the incident photon, which is ignored in the analytic approximation for simplicity.
Figure \ref{slafStaticSphereTest} (inset) shows the output of \textsc{slaf} with the `recoil effect' disabled, and demonstrates almost perfect agreement with the analytic solution.

\begin{figure}
\includegraphics[width=84mm]{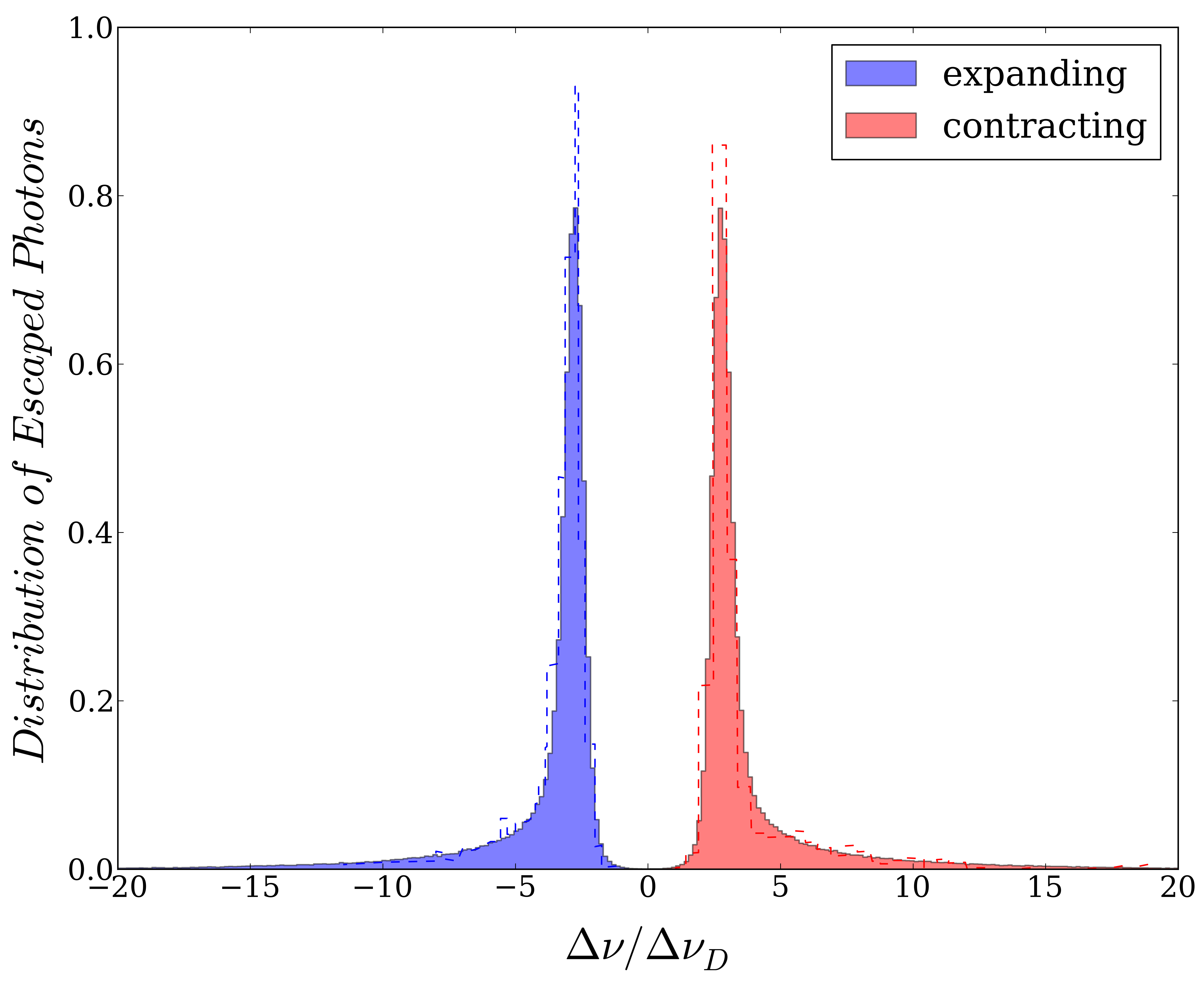}
\caption{Comparison of the spectrum calculated by \textsc{slaf} (histograms) with the monte carlo code (dashed lines) from \citet{2006ApJ...645..792T}.  Blue histograms/lines show a homogeneous expanding gas sphere, red histograms/lines show a homogeneous contracting gas sphere.  This plot simulates an expanding/contracting gas sphere with line-centre optical depth, $\tau=8 \times 10^4$ and $T=2 \times 10^4$~K.
$\Delta_{\nu}$ and $\Delta_{\nu_D}$ are as defined in Figure \ref{slafStaticSphereTest}}
\label{slafDynamicSpheresTest}
\end{figure}

To test that \textsc{slaf} can also correctly handle dynamic scenarios we make a comparison to the results presented in \citet{2006ApJ...645..792T} (Figure 3, right panel).
Since we want to make a direct comparison, we set up an identical scenario comprising of a single central \lya source embedded in a spherically symmetric HI cloud.

The HI cloud has a column density of $2 \times 10^{18}$ cm$^{-2}$ from the centre to the edge of the cloud, and a Hubble-like velocity gradient where the gas velocity scales proportionally to the radius up to a maximum inflow/outflow velocity of 200~km~s$^{-1}$ at the edge of the cloud.

Figure \ref{slafDynamicSpheresTest} shows our results overlaid on the results from \citet{2006ApJ...645..792T}.
The results in blue (red) are for the expanding (contracting) cloud case.
The solid histograms show our \textsc{slaf} results and the dashed-line histograms show the \citet{2006ApJ...645..792T} results.
There is excellent agreement between the two codes.

\subsection{Absorption along Sightlines (LAF)}

\textsc{laf} (Lyman Alpha Fighter) is another new code which calculates \lya absorption line profiles for sightlines through 3D volumes of arbitrary gas distributions and associated velocity fields.
Equivalent widths for sightlines can then be derived from the line profiles.

In order to test \textsc{laf} we construct a simple test case.
We create a sphere of HI gas with radius $r=25.6$~kpc, uniform HI number density $n=1\times10^{-10}$cm$^{-3}$, and a temperature of 5000~K on a 512$^3$ grid.

We then use \textsc{laf} to calculate the absorption line profiles, and hence absorption equivalent width, for lines of sight through the simulation volume.
For such a simple geometry the optical depth along lines of sight through the volume can be calculated analytically, and used to derive the equivalent width of the absorption lines along the line of sight.

\begin{figure}
\includegraphics[width=84mm]{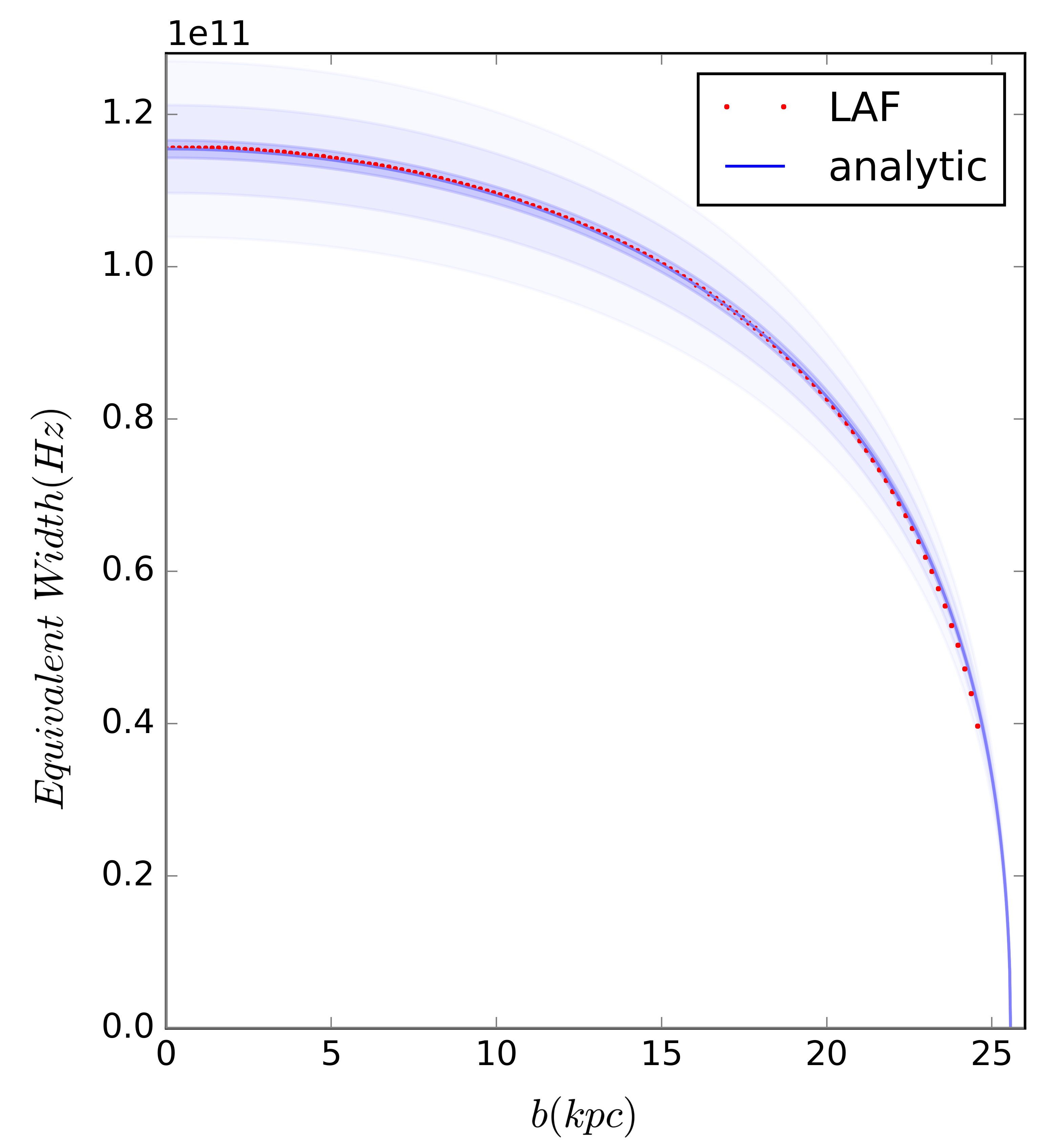}
\caption{Equivalent width vs impact parameter for a 25.6~kpc static, homogeneous HI sphere.  The blue line is the analytic solution while the shaded regions show 1\%, 5\% and 10\% errors.  The red dotted line shows the output from \textsc{laf} to be in excellent agreement with the analytic solution.}
\label{lafStaticSphereTest}
\end{figure}

Figure \ref{lafStaticSphereTest} shows a comparison between the \textsc{laf} output and the analytic solution.
The output of \textsc{laf} is in excellent agreement with the analytic solution, with a small deviation arising only from the fact that the sphere has been quantised onto a grid.
This can be seen in the fact that the agreement gets worse as the impact parameter approaches the radius of the sphere.
This can be understood by realizing that as a sightline moves close to the edge of the sphere the segment intersecting the gas sphere decreases.
Thus the relative error due to the discretisation onto a grid grows as the traversed optical depth decreases with higher impact parameter.

\subsection{Robustness to the chosen domination criterion}\label{dom_criteria}

There is scope for concern about the specific criterion chosen to ascertain whether a particular cell is dominated by the ionising radiation from local sources or from the UVB.
Our pipeline uses this criterion to decide whether a cell's ionisation state is calculated via `real' ionising radiative transfer using \textsc{crash} or assumed to be the UVB equilibrium ionisation state.
However the issue is not as critical as it may at first appear.
This is because the UVB is never fully ignored, even when a cell is determined to be local-source dominated.

To calculate the ionisation state of cells the ionisation equilibrium state with the UVB is first calculated for all grid cells.
This is used as the initial conditions for the next step, which is to add in local ionising radiation sources, and perform radiative transfer to calculate the new ionisation state of cells.
Because our radiative transfer simulation does not include the UVB, recombinations may occur during the radiative transfer of the local ionising photons which in reality should not (due to the presence of the UVB).
This is why we test whether each cell is dominated by the UVB or by local sources.
If a cell is UVB dominated, we use the UVB equilibrium state, otherwise we use the result of our local-source radiative transfer.
However, this local-source dominated ionisation state already implicitly includes the UVB in the initial conditions.

Robustness to the domination criterion was also tested explicitly.  
To do this, the criterion used to decide whether a cell is local-source or UVB-dominated was adjusted.
In the main paper, a cell is deemed to be dominated by the UVB if the ionising flux from the UVB is greater than the ionising flux from local sources.
We modified the criterion so that a cell is considered to be UVB dominated if the ionising flux is 2 or 3 times higher than the local flux.
This serves to inform us how sensitive our procedure is to this criterion.

\begin{figure}
\includegraphics[width=84mm]{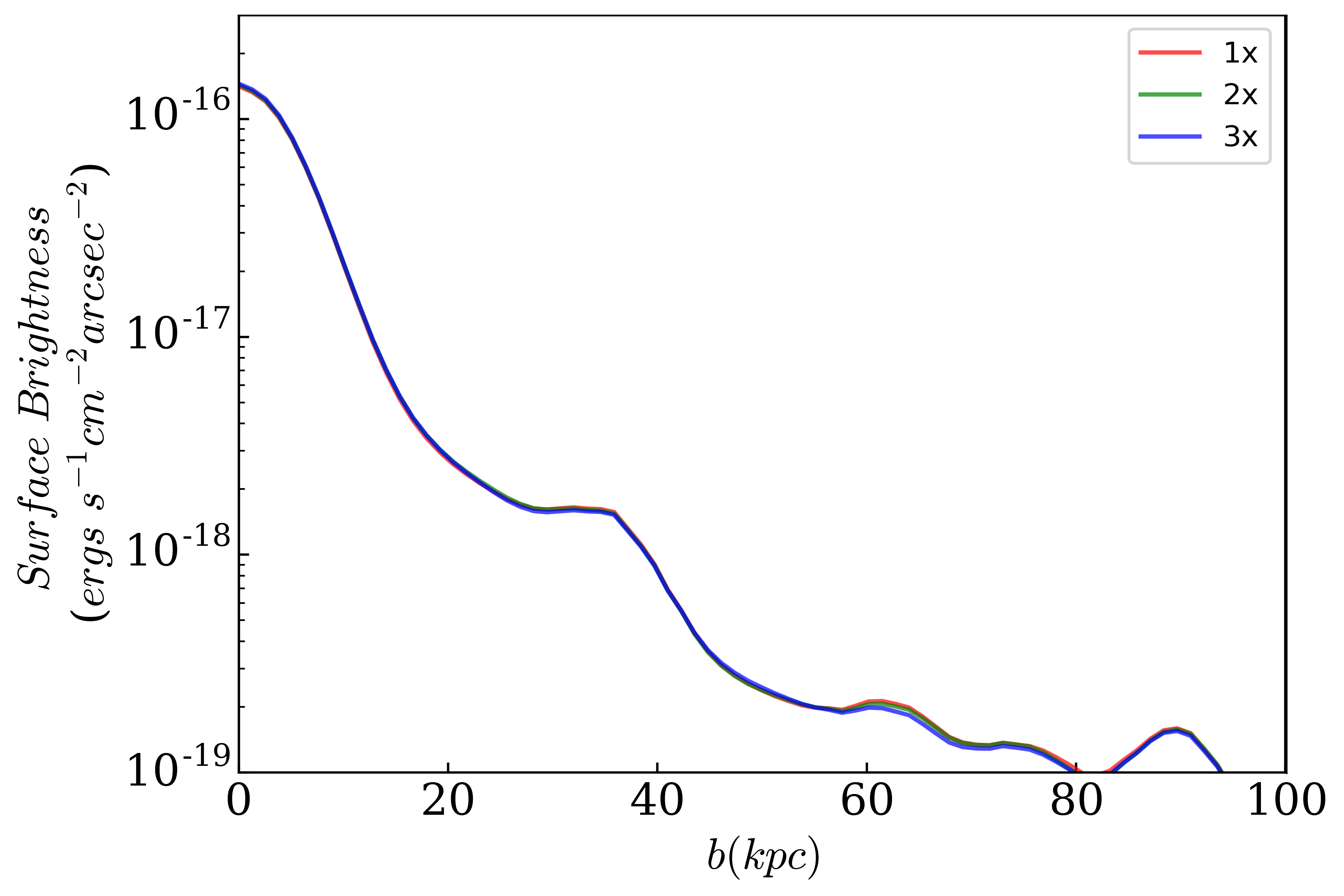}
\caption{Surface brightness profile as per Figure \ref{fig4} but for a single viewing orientation only.
The red, green, and blue lines show the results of using a 1x, 2x, and 3x UVB domination criterion respectively (see text).
The profiles deviate by up to $\sim 15\%$.}
\label{dom_criteria_plot}
\end{figure}

The results (Figure \ref{dom_criteria_plot}) show that altering the domination criterion affects the surface brightness profile by up to $\sim 15\%$ at any given impact parameter (and much less towards the centre of the galaxy).
On the logarithmic scale at which the data are presented, and the trends manifest themselves, this $\pm 15\%$ variation is so small as to be irrelevant.

\subsection{On ignoring \lya from recombination}\label{appendix_recombination}
As per \S~\ref{emission}, \lya emission resulting from recombinations in the CGM is ignored in our simulations.
For brevity this is justified in the main text by noting that recombination in the CGM accounts for less than 10\% of the total \lya emission.
However, when deciding whether recombination emission can be ignored, we must also consider the spatial distribution of the emission.
If the spatial distributions of the two sources differ significantly, and in such a way that the recombination emission becomes dominant at certain impact parameters, then radiative transfer effects could potentially alter the resultant surface brightness profile.

\begin{figure}
\includegraphics[width=84mm]{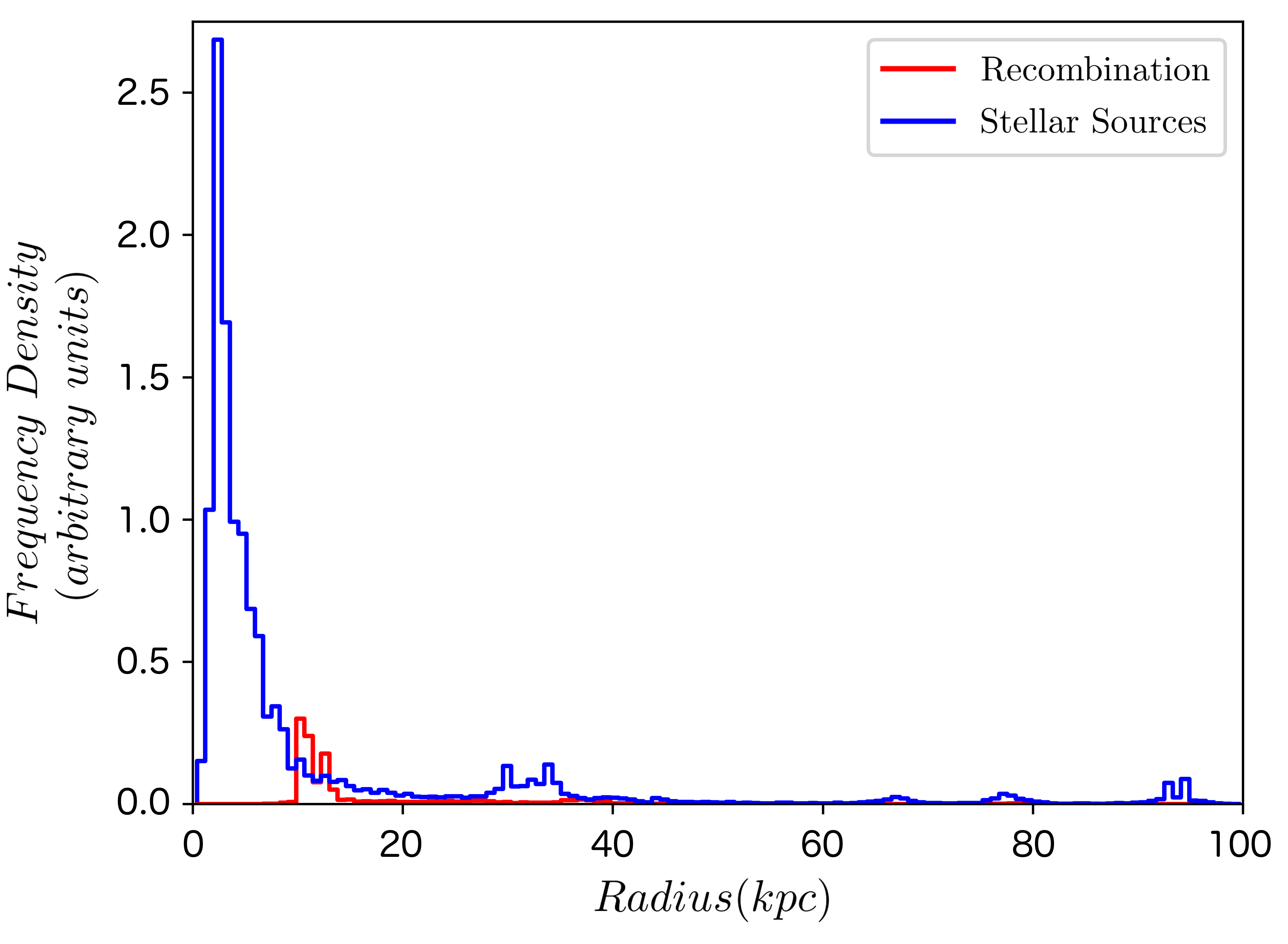}
\caption{A comparison of the radial distribution of stellar emission and CGM recombination emission, where recombination emission is set to 10\% of the stellar emission.
The emission is in arbitrary units but normalised such that the two plots are directly comparable.
For a closer crop see Figure \ref{spatial_distribution_crop}.}
\label{spatial_distribution}
\end{figure}

\begin{figure}
\includegraphics[width=84mm]{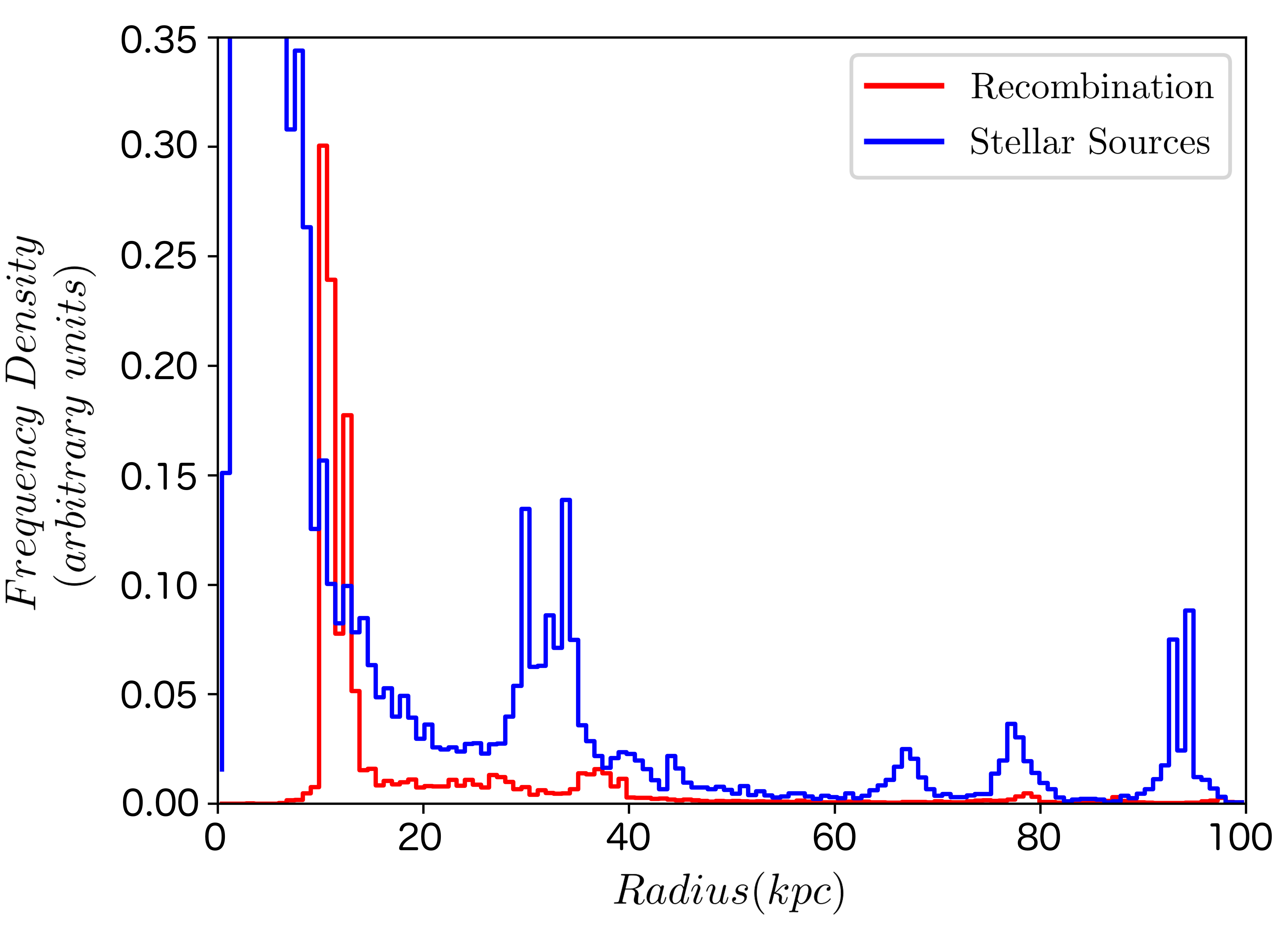}
\caption{Cropped detail view of Figure \ref{spatial_distribution}.}
\label{spatial_distribution_crop}
\end{figure}

We directly test for this by estimating the \lya recombination emission per cell of the simulation.
This assumes Case A or Case B recombination, depending on the physical state of each cell.
The radial distribution of the recombination emission is then compared to the radial distribution of the stellar emission (Figure \ref{spatial_distribution} and Figure \ref{spatial_distribution_crop}).

Figure \ref{spatial_distribution} shows that the distribution of the recombination emission is strongly peaked around the galaxy, which is also the dominant source of \lya from stellar populations.
The peak of the recombination emission is offset from the centre of the galaxy.
This is due to the fact that in our simulations the very high density ISM gas is replaced by vacuum and a \lya escape fraction(see \S~\ref{ionisingRT}).
Therefore the highest amount of recombinations occur at the interface of this (artificial) vacuum region.
Because of this offset the plot appears (falsely) to show that the recombination emission dominates at a galactocentric radius of $\sim$8 kpc.
However, the vacuum which causes the recombination emission to be peaked offset from the centre of the galaxy, also causes the stellar emission to have a similar offset.
The \lya photon packets from the stellar sources in the vacuum region free-stream until they hit this interface, and so the \emph{effective} position of the sources (and all of the emission in the central $\sim$8kpc in Figure \ref{spatial_distribution} and Figure \ref{spatial_distribution_crop}) is also at this interface.
This is not manifested in Figure \ref{spatial_distribution} and Figure \ref{spatial_distribution_crop} because they show the spatial distribution of the sources, ignoring radiative transfer effects.

This quirk of our simulations aside, it is clear from Figure \ref{spatial_distribution_crop} that the stellar sources dominate at all galactocentric radii.
Due to the vagaries of \lya radiative transfer, we cannot \emph{guarantee} that the CGM recombination emission is negligible without explicitly including it in the radiative transfer.
Nevertheless, we hope that the above discussion gives some insight into why, on balance, we deem it highly probable that \lya from CGM recombination can safely be ignored with respect to the main results of this paper.

\bibliographystyle{mn2e}
\bibliography{refs}
\label{lastpage}
\end{document}